# Effect of Hydrostatic Pressure on Superconductivity in κ-$[(BEDT\text{-}TTF)_{1-X}(BEDSe\text{-}TTF)_X]_2Cu[N(CN)_2]Br$

**Y. V. Sushko[1], S. O. Leontsev[1], O. B. Korneta[1], A. Kawamoto[2]**

[1] *Department of Physics and Astronomy, University of Kentucky, Lexington KY 40506, USA*

[2] *Division of Physics, Hokkaido University, Sapporo 060-0810, Japan*

**Abstract**. Static susceptibility of κ-$[(BEDT\text{-}TTF)_{1-x}(BEDSe\text{-}TTF)_x]_2Cu[N(CN)_2]Br$ alloys with the BEDSe-TTF content near the border-line of ambient pressure superconductivity (x~0.3) has been measured as a function of temperature, magnetic field, and pressure. A non-monotonic pressure dependence is observed for both the superconducting critical temperature and superconducting volume fraction, with both quantities showing growth under pressure in the initial pressure range P < 0.3 kbar. The results are discussed in comparison with the data on the related kappa-phase BEDT-TTF superconductors in which not a cation but anion sublattice is modified by alloying, namely the family κ-$(BEDT\text{-}TTF)_2Cu[N(CN)_2]Cl_{1-x}Br_x$.

## 1. INTRODUCTION

Recently, Sasaki et al.[1] reported synthesis of a new member of the family of kappa-phase (BEDT-TTF) superconductors, namely a series κ-$[(BEDT\text{-}TTF)_{1-x}(BEDSe\text{-}TTF)_x]_2Cu[N(CN)_2]Br$ in which organic cation layers are composed of a mixture of (BEDT-TTF) and (BEDSe-TTF) molecules. The prototype $x$ = 0 salt, an ambient pressure superconductor[2] with $T_c$ = 11.5K as well as an isostructural compound, the Cl-salt, known to be a weak ferromagnet[3] and insulator at ambient pressure, remain a subject of intensive theoretical and experimental investigations since their discovery more than 15 years ago. In 1998 the $x = 1$ salt κ-(BEDSe-TTF)$_2$Cu[N(CN)$_2$]Br was reported by Sakata et al.[4] to be a metal undergoing transition to an insulating and magnetic (supposedly SDW) state below ~25K and becoming superconducting only under an applied pressure of more than 1.5 kbar (with a $T_c$ of ~7.5K, about 30% smaller than in the BEDT-TTF analog). Interesting findings of the work of Sasaki et al. on the alloys are that (i) superconducting state exists even at ambient pressure, provided that (BEDSe-TTF) concentration is relatively weak, $x$<1/3, and (ii) the dependence of $T_c$ on $x$ is approximately monotonic.

Extreme sensitivity to an applied pressure is a well known property of the BEDT-TTF kappa-phase salts, leading, as in a case of κ-[(BEDT-TTF)]$_2$Cu[N(CN)$_2$]Cr, to a very complex pressure-temperature phase diagram[5], that exhibits, among other things, the phase coexistence (between a 13K superconducting state and a magnetic insulator state) with strong cooling-speed- and magnetic-field-history dependencies[6]. Since pressure effect data on materials with a mixed (BEDT-TTF)/(BEDSe-TTF) cation sublattice are not available yet, we studied and report here the magnetization of κ-[(BEDT-TTF)$_{1-x}$(BEDSe-TTF)$_x$]$_2$Cu[N(CN)$_2$]Br alloys under pressure.

## 2. EXPERIMENTAL

The single crystals with $x = 0.17$ and $x = 0.27$, grown by a standard electrocrystallization technique[1], were studied. Susceptibility measurements under pressure were performed with a Quantum Design MPMS SQUID magnetometer combined with a helium-gas pressure technique which provides good strain homogeneity through the entire sample volume[7]. In all reported here experiments a measuring magnetic field was applied normal to the **ac** plane of a crystal. The data are not corrected for the demagnetization factor.

## 3. RESULTS AND DISCUSSION

The susceptibility vs. temperature plot shown in Figure 1 presents our finding that a magnitude of diamagnetic signal in the alloy compound strongly depends on the cooling rate. Namely, for the x = 0.17 salt, cooling from 300K with a rate of 0.5K/min results in a diamagnetic signal that for both zero field cooling (ZFC) and field cooling (FC) processes is approximately 50 times stronger than after cooling at a rate of 5K/min. For the x = 0.27 salt, a rate of 5K/min suppresses superconductivity completely.

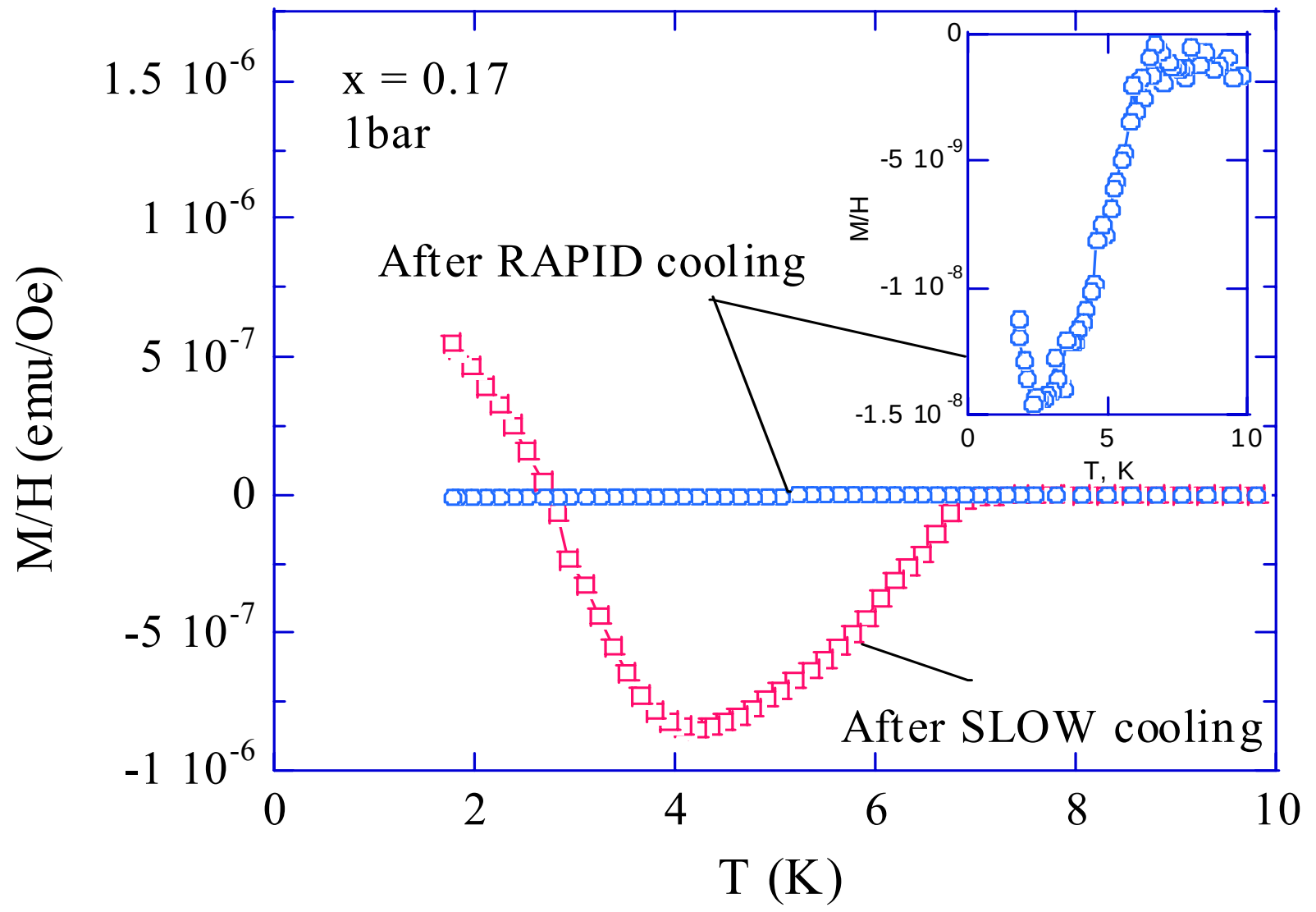

Fig. 1. M/H vs. temperature plots for *x*= 0.17 salt after slow and rapid cooling at P =1bar for a FC process.

We also note that a shape of the field cooled (FC) magnetization vs. temperature curve shown in Fig.1 is rather peculiar, with a largest diamagnetic signal observed not at the lowest temperature, but at ~4K, and with an appearance and growth of a paramagnetic signal under further lowering temperature. No such effect was observed in ZFC magnetization.

As shown in detail in Figure 2, an increase in external magnetic field steadily suppresses the reentrance of paramagnetism, and for H > 150 Oe the change of magnetization with temperature becomes monotonic.

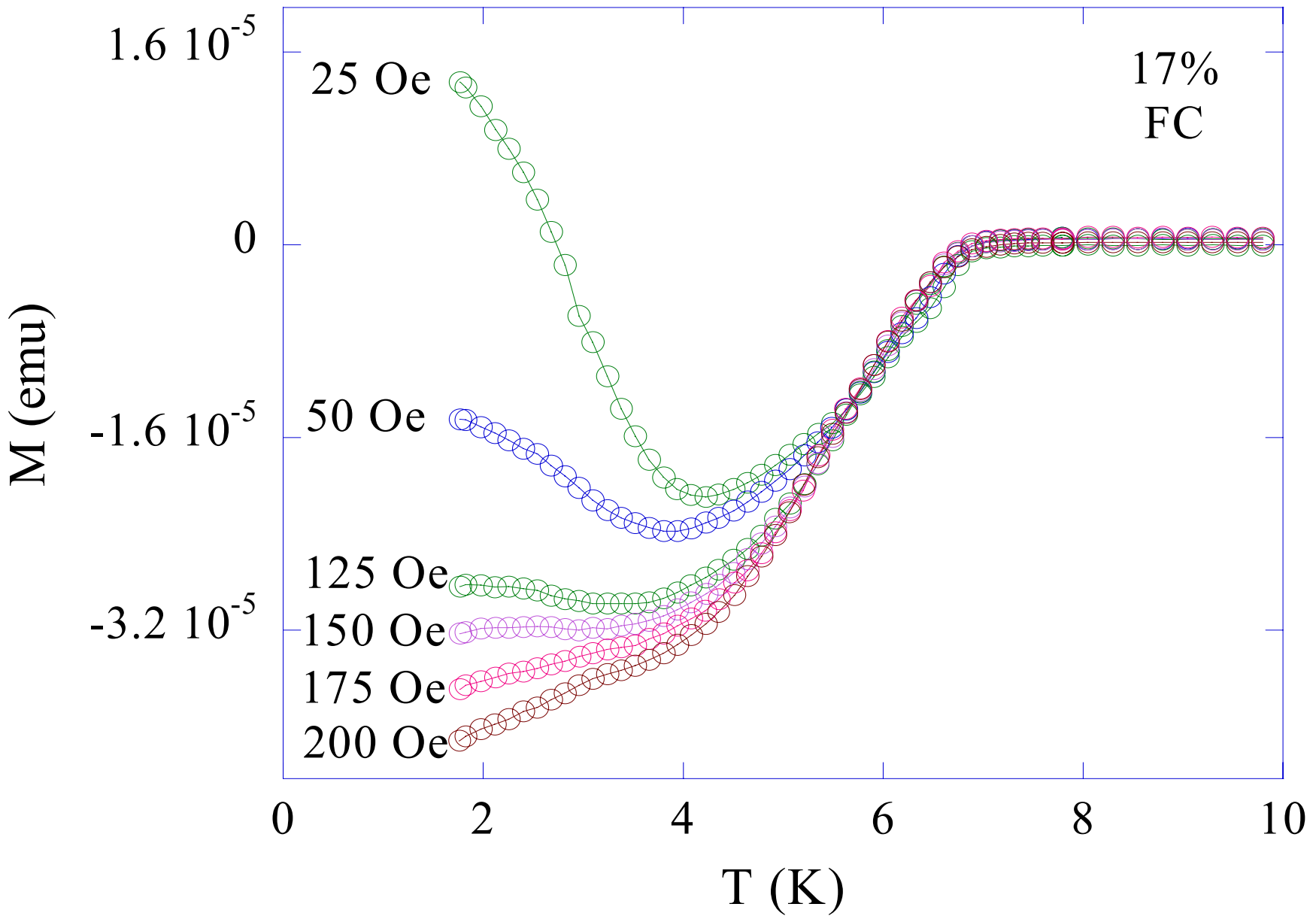

Fig. 2. FC branch of M vs. T dependence for $x = 0.17$ salt in different magnetic fields.

For the $x = 0.27$ salt no reentrance of paramagnetism was observed, as illustrated by the shown in Fig. 3 M(T) curves measured on cooling in fields of 10 and 100 Oe.

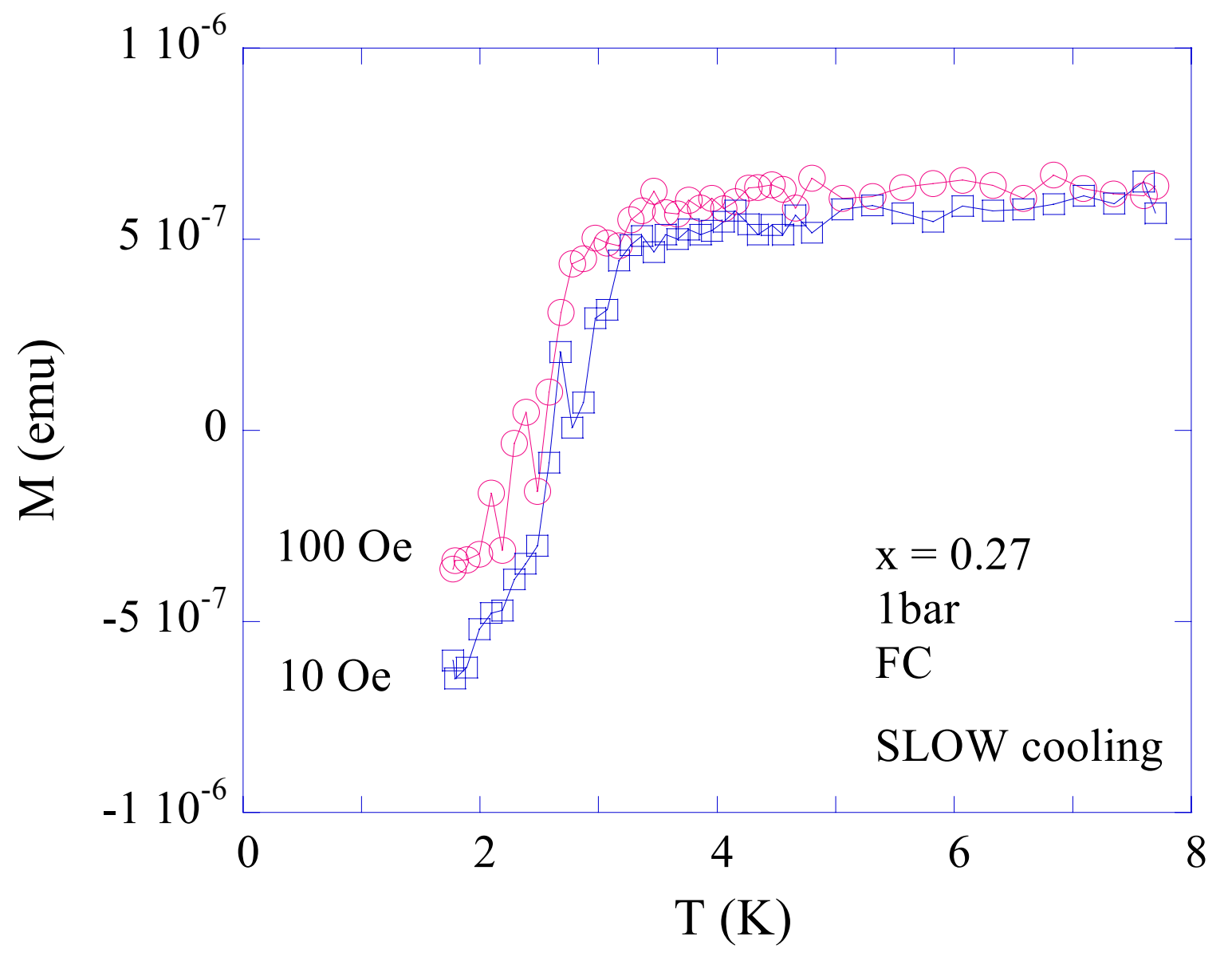

Fig. 3. FC branch of M vs. T dependence for $x = 0.27$ salt in different magnetic fields.

The effect of pressure on the shielding effect in the $x = 0.17$ samples is presented in Figure 4. The main feature is a shift of superconducting transition to lower temperatures with a rate $dT_c/dP = -2.7K/kbar$.

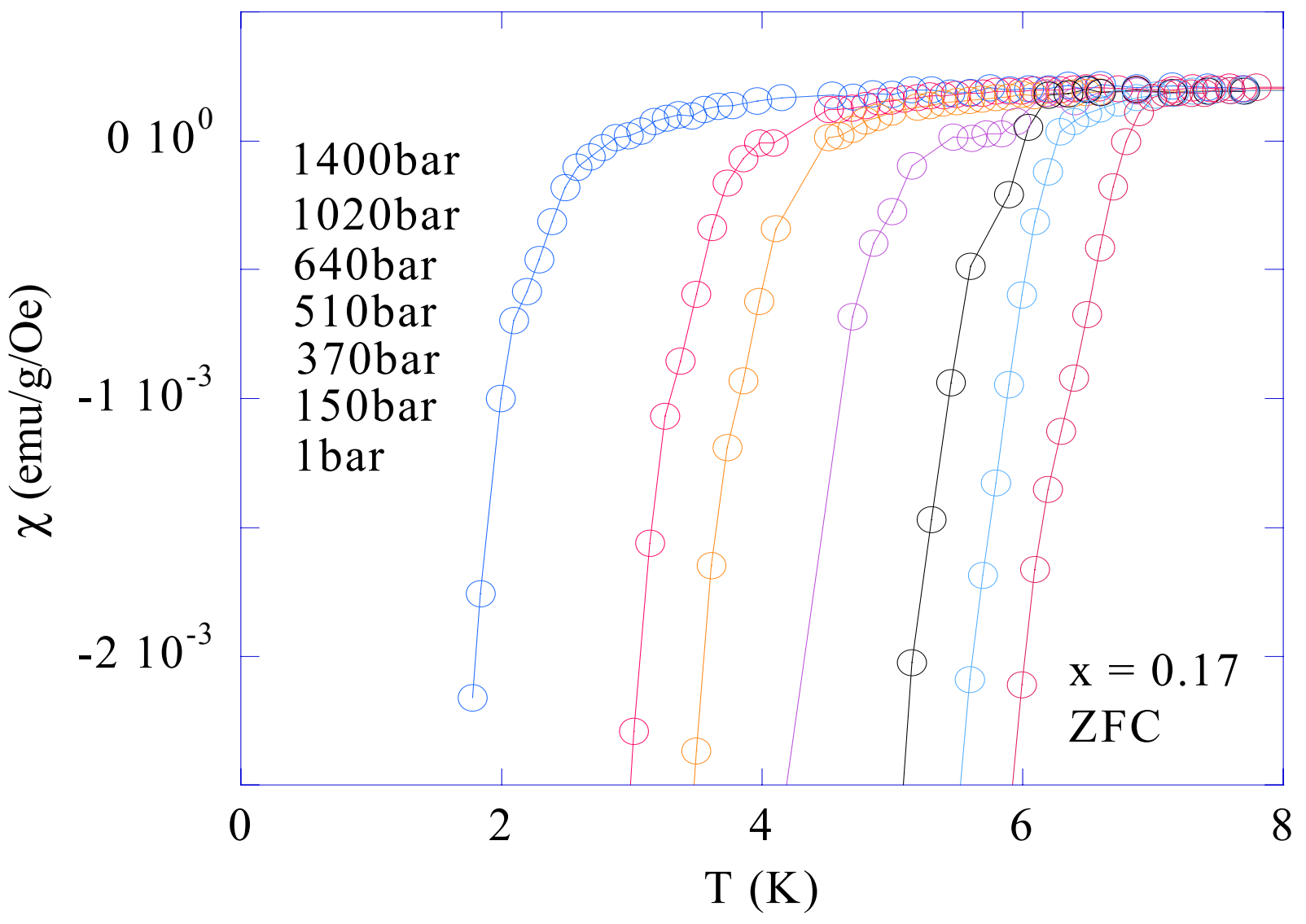

Fig. 4. Temperature dependence of a ZFC susceptibility of the $x = 0.17$ salt at various pressures.

The pressure effect on superconductivity in $x$ = 0.27 salt is displayed in Figure 5. In a range P < 220-290bar, we observe an increase of $T_c$ with pressure, whereas for P>290bar the $T_c$ drops when pressure increases.

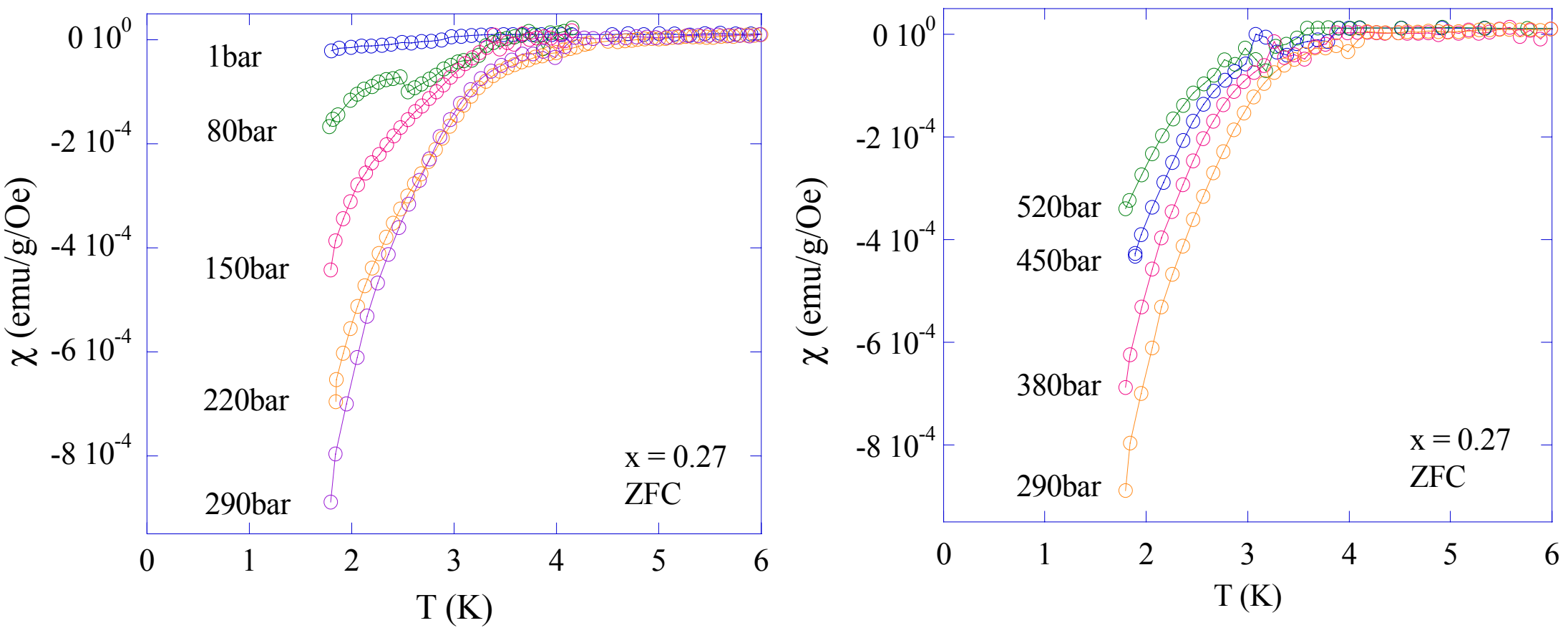

Fig. 5. Temperature dependence of the diamagnetic shielding of the $x$ = 0.27 salt at P ≤ 290bar (left panel) and P ≥ 290bar (right panel).

The results are summarized in Figure 6, where the $T_c$ *(P)* dependence for the $x$ = 0.17 salt is also shown. Remarkably, the data reveal that the rate of suppression of $T_c$ with pressure in the 3 K superconductor ($x$ = 0.27) is the same as in the 7 K superconductor ($x$ = 0.17), and also the same as in the BEDT-TTF salts[5,8] possessing the values of $T_c$ as high as 11.5-13 K. Yet another similarity with the prototype kappa-(BEDT-TTF) salts is underlined by the fact that the $T_c$ *(P)* dependence for the *x* = 0.27 alloy is non-monotonic with a clear maximum at pressures near 0.3 kbar.

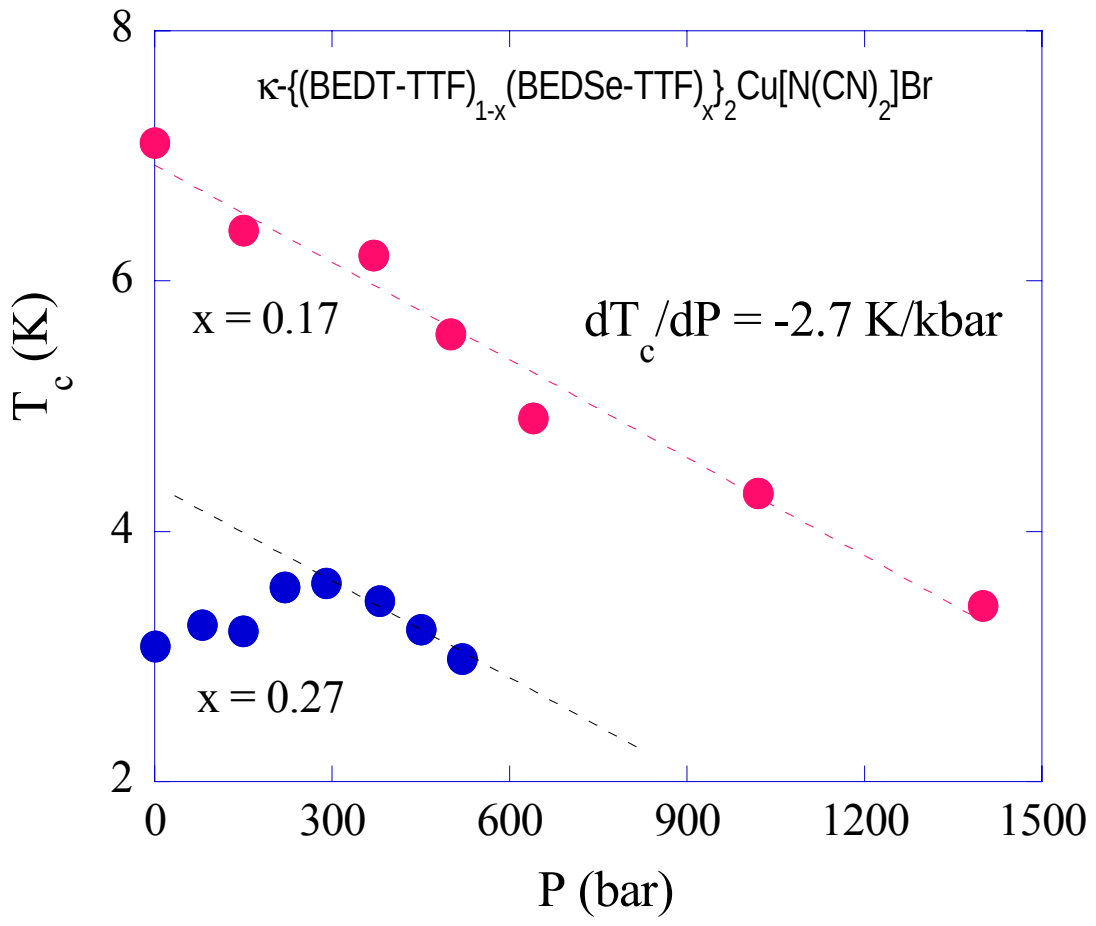

Fig. 6. Pressure dependence of the superconducting $T_c$ for the $x$ = 0.17 and $x$ = 0.27 compounds

Pressure dependence of the magnitude of FC diamagnetic susceptibility at 1.8 K in the *x* = 0.27 sample is displayed in Fig. 7. The data reveal that the Meissner effect of the *x* = 0.27 κ-[(BEDT-TTF)$_{1-x}$(BEDSe-TTF)$_x$]$_2$Cu[N(CN)$_2$]Br alloy gradually increases under pressure until the latter reaches a value of ~0.3 kbar - the same critical pressure that stabilizes an optimal superconducting volume fraction in κ-(BEDT-TTF)$_2$Cu[N(CN)$_2$] Cl$_{1-x}$Br$_x$ alloys with low (x<0.3) Br concentration ( see data of Figs. 2 and 3 of Ref. 8) . We also remark that for the *x* = 0.27 salt the superconducting volume, even when enhanced by an external pressure P>0.3 kbar, remains smaller than in the *x* = 0.17 salt by an order of magnitude, and by an additional factor of 3 if compared to the pure kappa-(BEDT-TTF) salt.

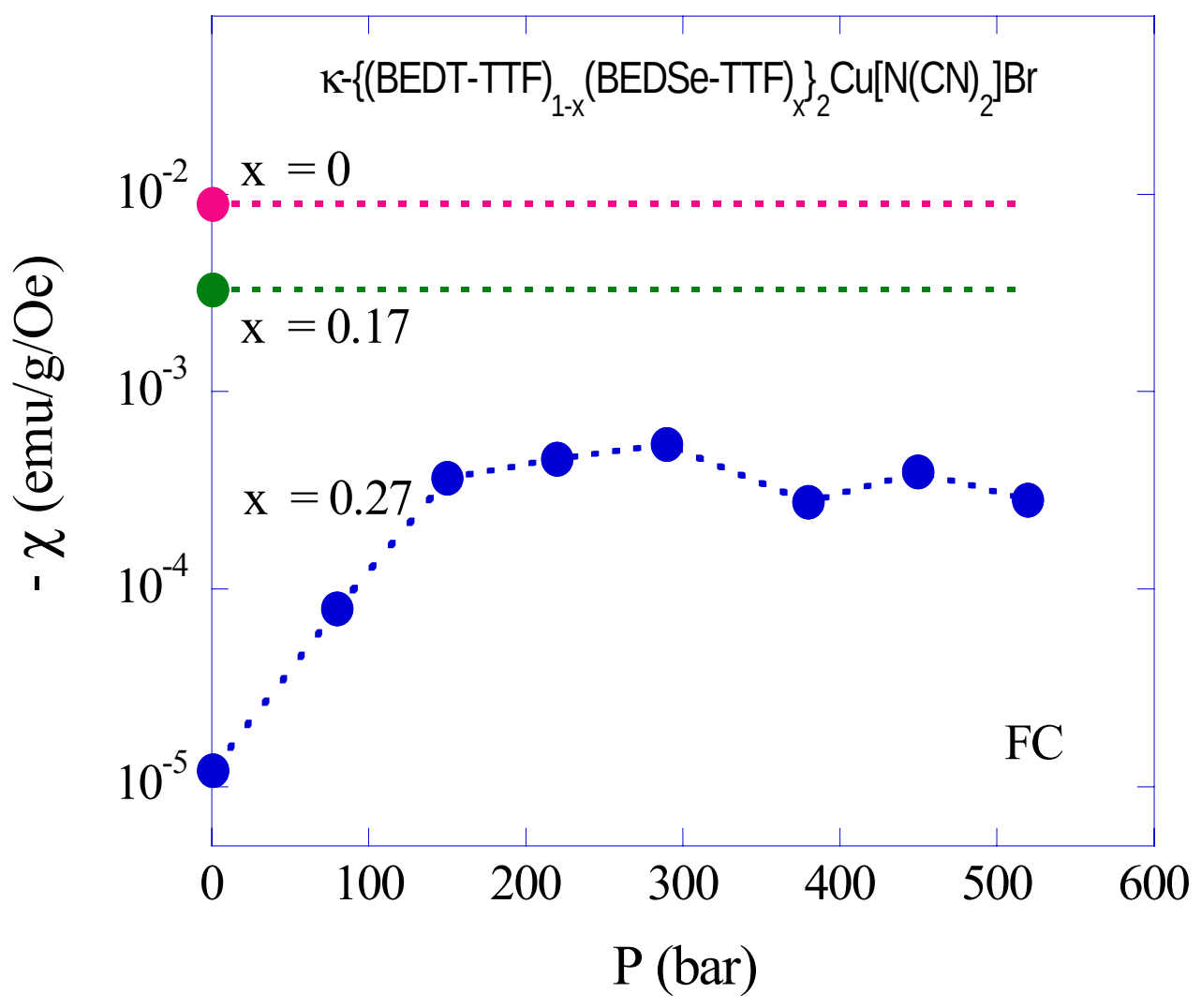

Fig. 7. Pressure dependence of the Meissner expulsion at 1.8 K in the *x* = 0.27 salt. Data for the *x* = 0 and 0.17 salts are shown for comparison.

## CONCLUSIONS

In conclusion, we have measured the dc magnetization of single crystals of two -[(BEDT-TTF)$_{1-x}$(BEDSe-TTF)$_x$]$_2$Cu[N(CN)$_2$]Br compounds, with $x$~0.17 and $x$~0.27, using a SQUID magnetometer combined with a helium-gas hydrostatic pressure technique. For the BEDSe-TTF content $x$~0.27, which is on the border-line of ambient pressure superconductivity, a non-monotonic pressure dependence is observed for the superconducting critical temperature as well as superconducting volume fraction, with both quantities showing growth under pressure in the initial pressure range $P < 0.3$ kbar. For $P > 0.3$ kbar, the $T_c$ of this compound decreases at the same rate of 2.7K/kbar as for the $x = 0.17$ salt ( although in the latter case $T_c$ decreases monotonically in the whole pressure range $P > 0$). The results appeared to be strikingly similar to the behaviour previously observed in the kappa-phase BEDT-TTF superconductors in which not a cation but an anion sublattice has been modified by alloying, namely the -(BEDT-TTF)$_2$Cu[N(CN)$_2$]Cl$_{1-x}$Br$_x$ family.

The support from NSF (Grant No. DMR 05-02706) is acknowledged.